\documentclass[twocolumn]{aastex62}

\newcommand{\Msun}{$M_{\odot}$}

\newcommand{\el}[2]{$^{#2}$#1}

\received{October 27, 2019}
\revised{November 18, 2020}
\accepted{November, 19, 2020}
\submitjournal{ApJ}

\shorttitle{Nebular Sub-Chandras}
\shortauthors{Polin et al.}

\begin{document}

\title{Nebular Models of Sub-Chandrasekhar Mass Type Ia Supernovae: Clues to the Origin of Ca-rich Transients}

\correspondingauthor{Abigail Polin}
\email{abigail@berkeley.edu}

\author{Abigail Polin}
\affiliation{Department of Astronomy, University of California, Berkeley}
\affiliation{Department of Physics, University of California, Berkeley}
\affiliation{Lawrence Berkeley National Laboratory}

\author{Peter Nugent}
\affiliation{Department of Astronomy, University of California, Berkeley}
\affiliation{Lawrence Berkeley National Laboratory}

\author{Daniel Kasen}
\affiliation{Department of Physics, University of California, Berkeley}
\affiliation{Department of Astronomy, University of California, Berkeley}
\affiliation{Lawrence Berkeley National Laboratory}

\begin{abstract}
We use non-local thermal equilibrium (NLTE) radiative transport modeling to examine observational signatures of sub-Chandrasekhar mass double detonation explosions in the nebular phase. Results range from spectra that look like typical and subluminous Type Ia supernovae (SNe) for higher mass progenitors to spectra that look like Ca-rich transients for lower mass progenitors. This ignition mechanism produces an inherent relationship between emission features and the progenitor mass as the ratio of the nebular [\ion{Ca}{2}]/[\ion{Fe}{3}] emission lines increases with decreasing white dwarf (WD) mass. Examining the [\ion{Ca}{2}]/[\ion{Fe}{3}] nebular line ratio in a sample of observed SNe we find further evidence for the two distinct classes of SNe Ia identified in \cite{Polin2019} by their relationship between \ion{Si}{2} velocity and $B$-band magnitude, both at time of peak brightness. This suggests that SNe Ia arise from more than one progenitor channel, and provides an empirical method for classifying events based on their physical origin. Furthermore, we provide insight to the mysterious origin of Ca-rich transients. Low mass double detonation models with only a small mass fraction of Ca (1\%) produce nebular spectra that cool primarily through forbidden [\ion{Ca}{2}] emission. 
\end{abstract}


\section{Introduction}
\label{sec:intro}
A popular sub-Chandrasekhar model for Type Ia supernovae (SNe) consists of a carbon-oxygen white dwarf (C/O WD), below the Chandrasekhar mass limit, which ignites through the aid of an accreted helium shell. The helium ignition sends a shock wave into the center of the C/O WD causing a detonation near the core of the star followed by thermonuclear runaway. This process, known as the double detonation scenario, has been theorized as a possible path to SNe Type Ia, initially for WDs with large helium shells \citep{Woosley&Weaver94,Nomoto1982b,Nomoto1982a,Livne1990} and later for WDs with only a small amount of helium on their surfaces \citep{Bildsten2007,Shen14,Fink2007,Fink2010,Sim2010}. Current simulations show that the latter can be a promising path to normal and subluminous SNe Ia \citep{Shen2018,Polin2019,Townsley2019}.

Evidence increasingly points toward sub-Chandra-sekhar mass WDs being responsible for a significant portion of SNe Type Ia. Recently the discovery of SN~2018byg (ZFT18aaqeasu) provided strong evidence for a sub-Chandrasekhar mass explosion triggered by a massive helium shell ignition \citep{De2019}. \cite{De2019} fit this peculiar Type I SN with a model derived from \cite{Polin2019} concluding that the early flux excess exhibited by SN~2018byg was due to the radioactive decay of elements in the outermost ejecta which were produced during the initial helium shell burning. The spectrum at peak was also completely line blanketed for wavelengths less than 5000~\AA~which was fit well by models with optically thick helium shell ashes in the outermost ejecta. This event lends credence to the ability of the double detonation scenario to lead to the explosion of a WD. 

Other rare transients may also relate to WDs accreting helium shells. Ca-rich gap transients, so called because their peak luminosity lives in the ``gap" between that of novae and SNe, are a class of astrophysical transients identified by their nebular spectra, which are dominated by [\ion{Ca}{2}] $\lambda\lambda$7291, 7323 emission \citep{Kasliwal2012}. The origin of these transients is still unknown, but helium shell ignitions have been suggested as a possible progenitor \citep{Dessart2015}. These events all exhibit a fast photospheric evolution, rising in $\sim$15 days, with photospheric velocities $\sim$6,000-10,000 km/s. They reach the nebular phase very quickly, in 1-3 months, indicating a small ejecta mass. Furthermore, Ca-rich transients tend to occur offset from their host galaxies, indicating an origin from an old stellar population \citep{Lunnan2017}.

While peculiar transients may arise from WDs with a large amount of helium on their surface, studies have shown that smaller helium shells allow for normal and subluminous Type Ia events \citep{Polin2019,Townsley2019}. \cite{Polin2019} points to a population of SNe Ia distinguished by the relationship between their peak luminosity and \ion{Si}{2} velocity as likely candidates for originating from a sub-Chandrasekhar mass double detonation, a delineation that was further supported by the examination of spectropolarimetry measurements of this population in \cite{Cikota19}. Furthermore, the new discovery of fast Gaia WDs have been pointed to as the potential surviving companions of a very thin helium shell ignition occurring during a dynamical merger of two WDs, known as the D6 mechanism \citep{Shen2018D6}.

While studies show promising results when comparing double detonation models to SNe Type Ia during the photospheric phase there has yet to be a comprehensive examination of what these models look like in the nebular phase when the ejecta becomes fully optically thin. This study aims to examine the observational consequences of the double detonation mechanism once the ejecta reaches the nebular phase. Using the ejecta profiles produced in \cite{Polin2019} we perform one-dimensional NLTE radiative transfer simulations to systematically examine the observational signatures of double detonations in the nebular phase.

In section \ref{sec:methods} we describe the methods used to model nebular spectra and detail the parameter space of models we explore. We examine the synthetic spectra and qualitatively compare them to the trends of Type Ia SNe in section \ref{sec:spectra}. In section \ref{sec:Ca} we discuss the strong [\ion{Ca}{2}] emission and discuss the consequences for the understanding of Ca-rich transients. We compare the models to an existing set of observed nebular SNe Type Ia in section \ref{sec:data} and discuss the overarching implications of this study in section \ref{sec:discussion}.

Alongside this paper, we have released the models created for this study. These are available online at
\href{https://github.com/aepolin/DoubleDetonationModels}{\url{https://github.com/aepolin/DoubleDetonationModels}}.

\section{Methods}
\label{sec:methods}
We begin with the homologous ejecta profiles produced in \cite{Polin2019} and use the NLTE radiative transport code developed for \cite{Janos2017} to produce synthetic spectra in the nebular phase.

\subsection{Initial Ejecta Profiles}

\begin{figure*}
    \centering
    \includegraphics[width=\columnwidth]{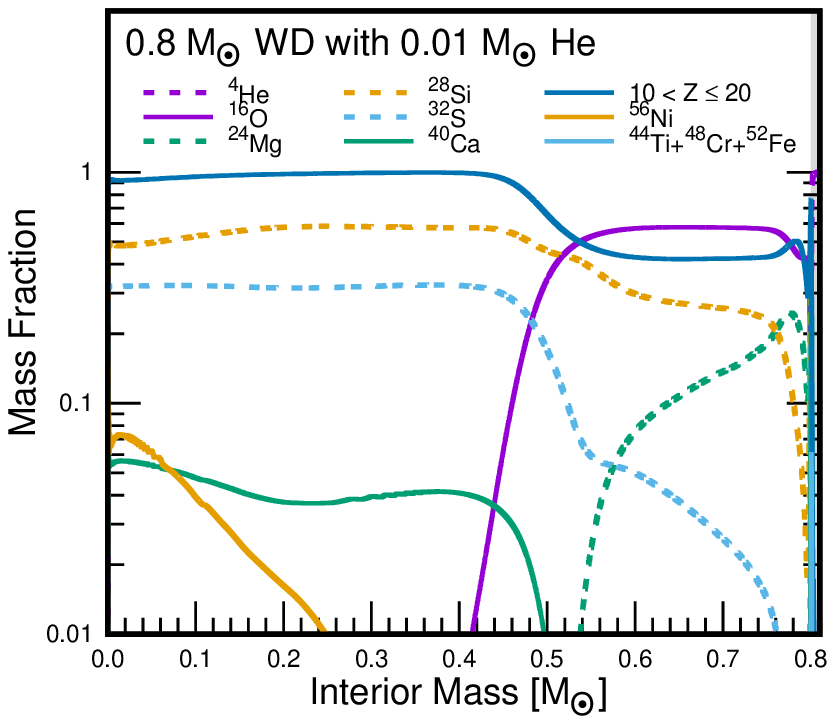}
    \includegraphics[width=\columnwidth]{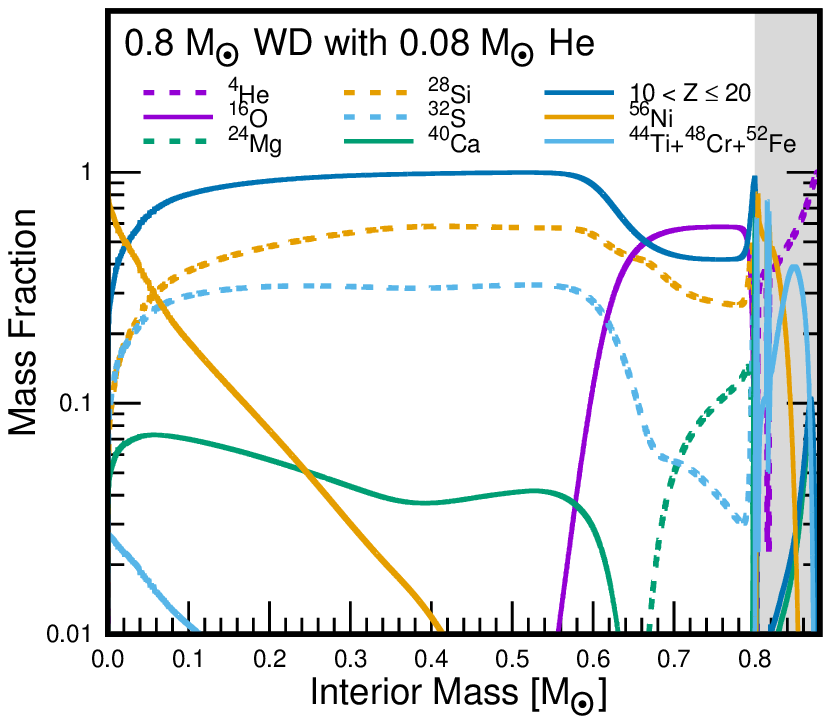}
    \includegraphics[width=\columnwidth]{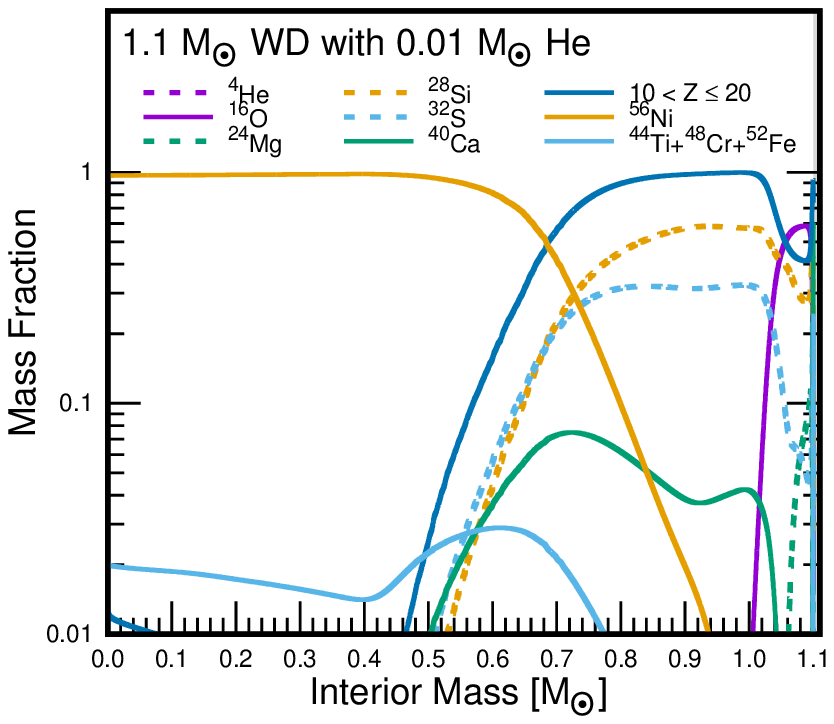}
    \includegraphics[width=\columnwidth]{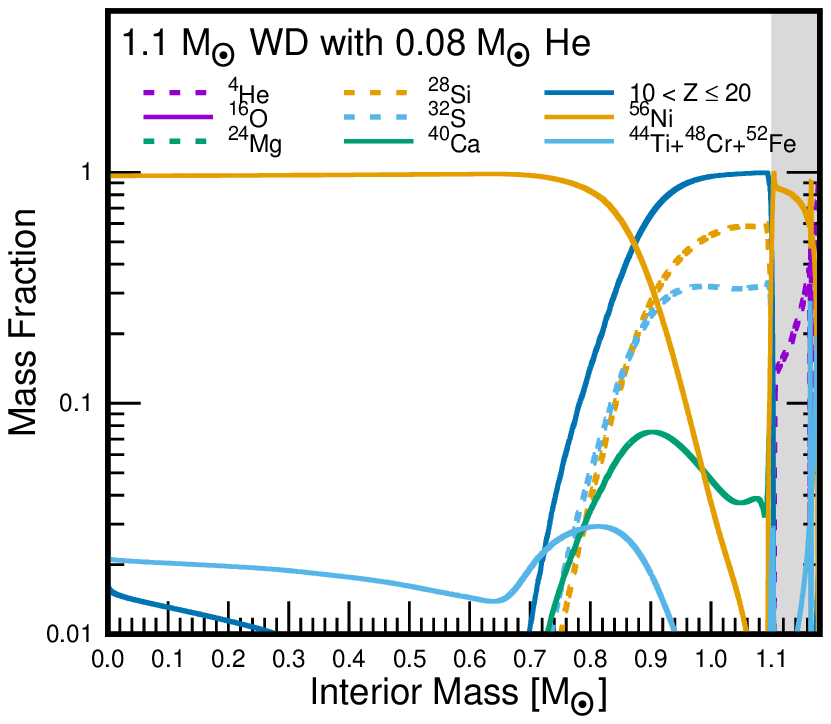}
    \caption{Example ejecta compositions from the \cite{Polin2019} models. We show examples of low mass (0.8 \Msun) WDs in the top row and high mass (1.1 \Msun) WDs on the bottom row, with thin helium shells (0.01 \Msun) on the left and thick helium shells (0.08 \Msun) on the right. The grey region in each plot denotes material that originates in the helium shell.}
    \label{fig:compositions}
\end{figure*}

\cite{Polin2019} explored a parameter space of double detonation models and their observational signatures in the photospheric phase. We begin with the \cite{Polin2019} 1D ejecta models that were produced using the hydrodynamics code CASTRO \citep{CASTRO}. Figure \ref{fig:compositions} shows a grid of the ejecta profiles of the double detonation models. We show a high mass WD (1.1 \Msun) as well as a low mass WD (0.8 \Msun) each with both a thick helium shell (0.08 \Msun) and a thin helium shell (0.01 \Msun). The core of these explosions resembles the profile expected for SNe Ia; a core of radioactive \el{Ni}{56}, surrounded by intermediate mass elements (IMEs), and lighter elements surrounding those. The precise composition created during the explosion depends on the mass and density profile of the WD, such that more massive (and denser) progenitors burn more completely and produce larger quantities of the heavier elements, including \el{Ni}{56}. Outside of the yields produced during the WD burning are the helium shell ashes. The composition of these ashes again depends on the initial progenitor; thick shells produce a significant amount of radioactive material in the outermost ejecta, which can cause early flux bumps in the photospheric light curves, while thinner shells only burn to IMEs and do not exhibit an early flux excess \citep{Polin2019}.

In this study we examine all of the \cite{Polin2019} models from 0.7-1.2 \Msun~WDs with 0.01-0.1 \Msun~He shells (35 models in total) and perform NLTE radiative transport calculations to produce synthetic spectra in the nebular phase. 

\subsection{NLTE Radiative Transport Methods}

\cite{Janos2017} have developed a 3D, NLTE, radiative transfer tool, \texttt{SedoNeb}, which we use to model the 1D nebular spectra presented in this work. The process involves first using the radiative transport code, \texttt{Sedona} \citep{sedona}, to model the gamma ray transport of radioactive decay products. We begin with the profile of each model once the ejecta has reached homologous expansion (when the velocity profile is no longer evolving with time, and is proportional to the radius) which occurs roughly a few seconds after the C/O WD ignition. We use \texttt{Sedona} to then transport the photons produced by the radioactive decay. \texttt{Sedona} includes contributions from the radioactive decay of  $^{56}$Ni$\rightarrow^{56}$Co$\rightarrow^{56}$Fe, $^{48}$Cr$\rightarrow^{48}$V$\rightarrow^{48}$Ti and $^{52}$Fe$\rightarrow^{52}$Mn$\rightarrow^{52}$Cr. From this we gain an understanding of the energy deposited throughout the ejecta as a function of time. We run our models through \texttt{Sedona} from 0.25 days until the nebular time of interest (150-500 days post explosion). At this epoch the primary source powering the ejecta is the radioactive decay of \el{Co}{56}, the daughter product of \el{Ni}{56} decay. While \texttt{Sedona} accounts for contributions to the energy deposition from other decay chains, which effect the observables of double detonations at early times \citep{Polin2019}, the half lives of \el{Vn}{48} ($\tau_{48Cr} \approx 1.3$ days, $\tau_{48Vn} \approx 23$ days) and \el{Mn}{52} ($\tau_{52Fe} \approx 0.5$ days, $\tau_{52Mn} \approx 0.02$ days) are significantly shorter than that of \el{Co}{56} ($\tau_{56Ni} \approx 8.8$ days, $\tau_{56Co} \approx 111$ days), and the amounts present are relatively small.

We then use \texttt{SedonNeb} to produce a nebular spectrum at a snapshot in time given the composition and energy deposition throughout the ejecta. With \texttt{SedoNeb} we calculate the emissivities of each atomic transition by solving for the temperature, ionization state, and NLTE level populations. Then we generate a spectrum by integrating the line emission to find the wavelength-dependent flux. \texttt{SedoNeb} considers emission contributions from H, C, O, Si, S, Ca, Fe, Co, and Ni, the primary species seen in nebular SNe Ia. Each spectrum presented is composed of 500 wavelength bins.

Some physical assumptions are made during this process. First, we assume that the gas temperature and level populations have reached equilibrium on a timescale much shorter than the expansion timescale. Next, we assume that by the epochs considered, the entire ejecta has become optically thin. This is a safe assumption for wavelengths of interest at epochs $\gtrsim$ 100 days after explosion, although the ejecta may remain optically thick in the ultraviolet for a longer time due to the contribution from iron-group elements, which remain opaque at these wavelengths for hundreds of days \citep{Friesen2017}.

\section{Nebular Spectra}
\label{sec:spectra}
In this section, we explore the spectral series produced by our simulations and discuss a qualitative comparison to the observed population of SNe Type Ia in the nebular phase.

\begin{figure*}
    \centering
    \includegraphics[width=\columnwidth]{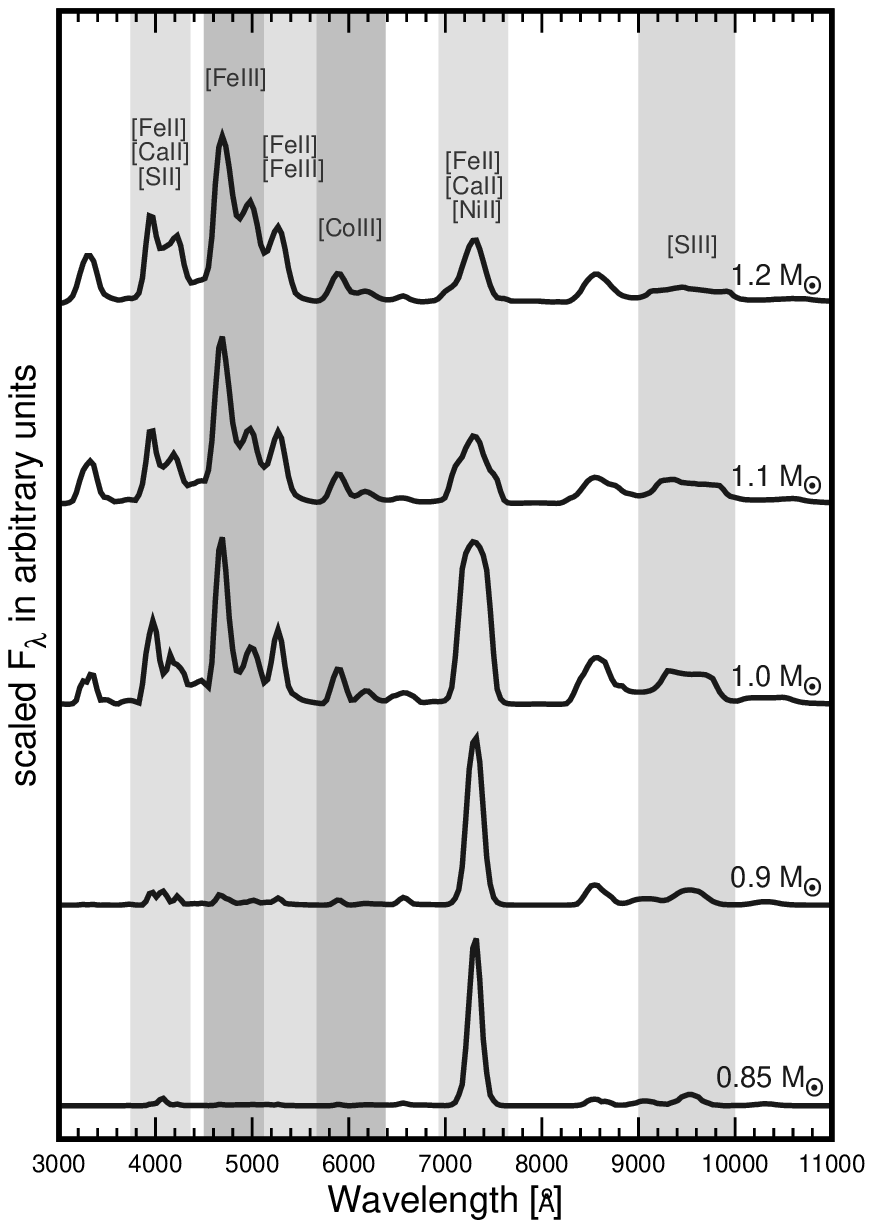}
    \includegraphics[width=\columnwidth]{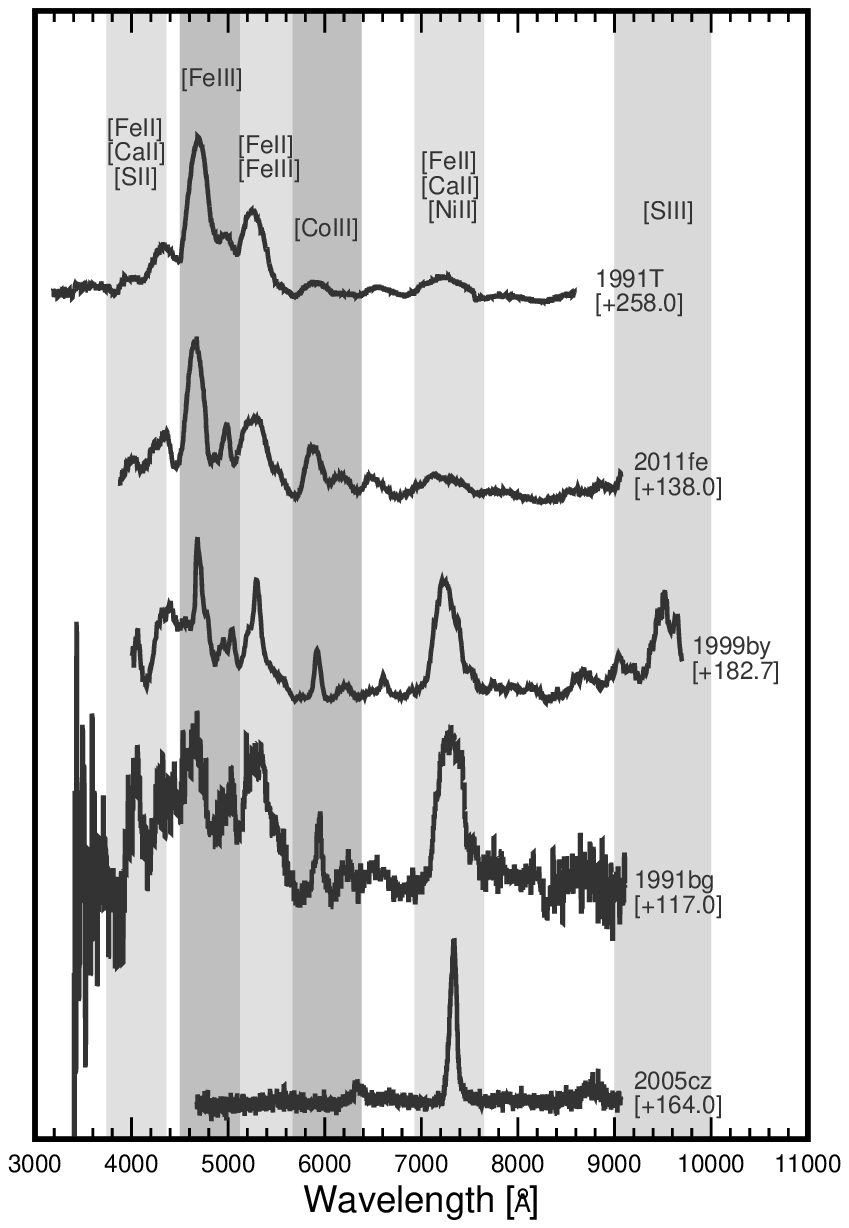}
    \caption{\textit{Left}: Nebular spectra of all models with 0.01 \Msun~He shells at 150 days after explosion. Models are arranged in decreasing mass from top to bottom, with labels on the right indicate the mass of the underlying WD for each. Strong [\ion{Ca}{2}] emission appears in all models. For models that produce less Ni-56 (and thus less [\ion{Fe}{2}] and [\ion{Fe}{3}] emission) the [\ion{Ca}{2}] $\lambda\lambda$7291, 7323 doublet is the predominant feature, reminiscent of Ca-rich transients. \textit{Right}: Observed sequence of nebular SNe Ia ordered from the most luminous (1991T) to the least luminous (1991bg) as well as an example Ca-rich transient (2005cz) at the bottom.}
    \label{fig:thinHeshellspectra}
\end{figure*}

\subsection{Summary of Results}

We take each model from \cite{Polin2019} and produce spectra for 150-450 days from explosion in increments of 50 days. 

The left panel of Figure \ref{fig:thinHeshellspectra} shows example spectra at 150 days from explosion, ranging from our most massive WD, 1.2 \Msun, at the top and decreasing in mass to a 0.85 \Msun~WD at the bottom, all with 0.01 \Msun~of helium on their surface. We first note that the spectra lack features that distinguish the mass of the helium shell. This result is not entirely surprising, as nebular spectra are probes of the internal structure of the SNe ejecta, and the inner ejecta of the double detonation models is primarily a function of the total mass of the progenitor ($M_{WD} + M_{He}$). For the rest of the paper, we focus on features that arise as we vary the total mass of the system. While we may not be able to use nebular spectra to determine the helium shell mass for a given double detonation explosion, the spectra produced by this progenitor channel likely differ from other sub-Chandrasekhar pathways to SNe Ia. For example, the collision of two WDs is expected to produce a bimodal velocity distribution, which would result in double peaked nebular features \citep{Dong2015}. 

Qualitatively the produced sequence reflects the trends seen in observed nebular SNe Type Ia (see the right panel of Figure \ref{fig:thinHeshellspectra}). The more massive progenitors produce spectra with strong Fe lines ($\sim$4500-5600~\AA), while also showing [\ion{Co}{3}] emission lines ($\sim$5800 - 6200~\AA). However, even our brightest model over produces [\ion{Ca}{2}] $\lambda\lambda$7291, 7323 emission when compared to the brightest SNe Ia (1991T and 2011fe). As we examine models with lower masses the spectra show increasingly weaker and narrower Fe emission until, for low enough mass progenitors, the majority of the cooling is through forbidden [\ion{Ca}{2}] emission. This, too, is a trend we see reproduced in the observed SNe Type Ia. Fe emission lines grow narrower for lower luminosity events \citep{Mazzali1998}, and the subluminous, 1991bg-like, SNe Ia show strong emission features around 7290~\AA~that can reach comparable strengths to their Fe emission (see 1991bg and 1999by in the right panel of Figure \ref{fig:thinHeshellspectra}). Our lowest mass models are reminiscent of the nebular spectrum of a Ca-rich transient. We further examine the implications of the [\ion{Ca}{2}] emission in section \ref{sec:Ca}.

\subsection{Normal Type Ia Supernovae}
\label{subsec:normal}

\begin{figure}
    \centering
    \includegraphics[width=\columnwidth]{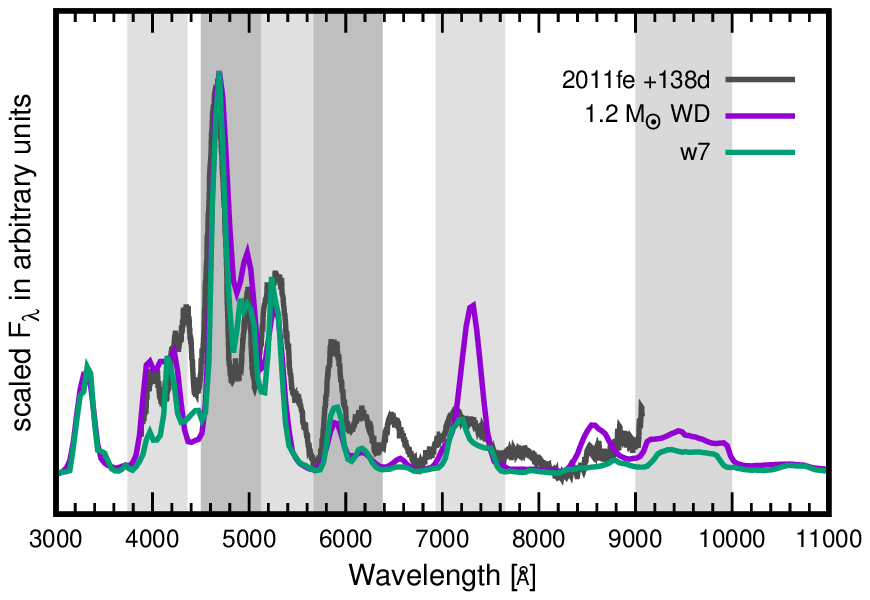}
    \includegraphics[width=\columnwidth]{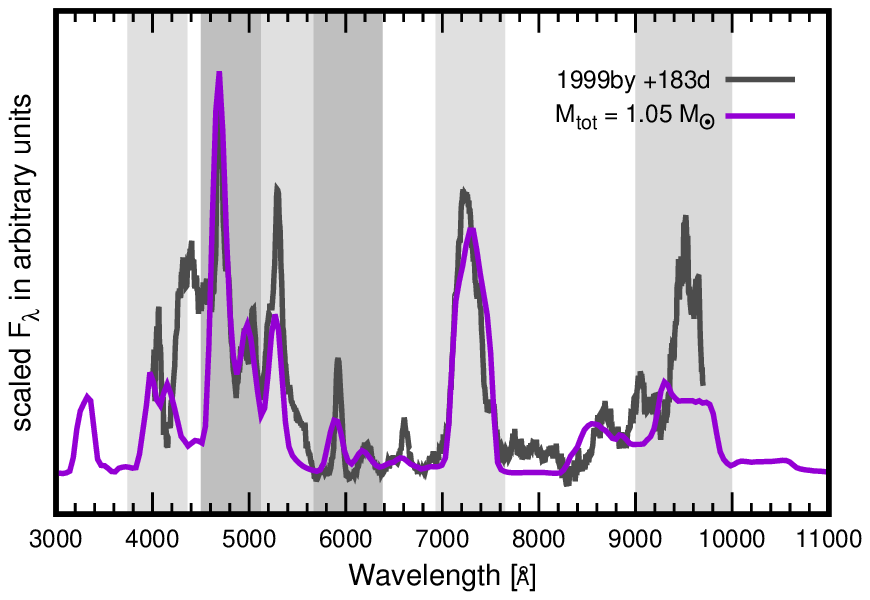}
    \caption{\textit{Top}: The normal Type Ia SN~2011fe at 138 days from peak \citep{2011fe_nebular} compared with our most luminous (and most massive) model (1.2 \Msun) in purple and the w7 model \citep{w7} in green. While our model produces a reasonable match to the strength and ratios of the Fe emission lines our model over produces Ca emission when compared to normal SNe Ia. The w7 model which contains far less $^{40}$Ca in its ejecta is a better fit. The w7 model ($M_V=-12.98$) is also a better match to the luminosity of SN~2011fe ($M_V=-13.37\pm0.05$) than our most massive model, which is less luminous ($M_V=-12.36$) at this phase. Distance to SN~2011fe taken from \cite{Maguire2014}.
    \\\textit{Bottom}: 1999by (a subluminous, 91bg-like Type Ia) at 183 days from peak \citep{Silverman2012} compared with our model for a progenitor with a total mass of 1.05 \Msun. Our model reproduces the strength of the emission feature at 7290~\AA. This model ($M_V=-10.88$) is also a good match to the luminosity of SN~1999by ($M_V=-10.95\pm0.2$) at this phase. Distance to SN~1999by taken from \cite{Garnavich2004}.}
    \label{fig:w711fe}
\end{figure}

In Figure \ref{fig:w711fe} we compare our most massive (and most luminous) model to the normal Type Ia SN~2011fe and the w7 pure deflagration model for Type Ia SNe \citep{w7}. At a glance the w7 model is a much better fit for a normal Type Ia SN than the double detonation model. While both models reproduce the strength and width of the Fe emission lines in the 4500-5500~\AA~range neither model perfectly reproduces the emission feature at 7290~\AA. This result is not surprising as the w7 model is a decent match to the peak and stretch of normal SNe Ia \citep{w7} while our double detonation models show some discrepancies with the observed properties in the photospheric phase \citep{Polin2019}.

Our results are mostly consistent with previous modeling of 2011fe performed in \cite{2011fe_nebular}. This study also notes an overproduction of [\ion{Ca}{2}] emission seen in nebular spectra produced by a sub-Chandrasekhar model in \cite{2011fe_nebular}. \cite{2011fe_nebular} propose successful model for the nebular signatures of 2011fe originating from a Chandrasekhar mass object with the inner most ejecta populated by stable Fe-group species. This stable Fe is required to reproduce the observed [\ion{Fe}{2}]/[\ion{Fe}{3}] ratio (as seen in the relative strengths of the 5270~\AA~and 4658~\AA~Fe emission lines respectively). The sub-Chandrasekhar mass model is further ruled out by its density profile, which would be less dense in the central regions during burning and produce very little stable Fe, an thus under produce [\ion{Fe}{2}] emission. While we agree that the [\ion{Ca}{2}] emission is in conflict with the sub-Chandrasekhar model for 2011fe it is important to note that our models do produce the relative strengths of the 5270~\AA~and 4658~\AA~lines without this stable Fe present in the ejecta.

The 7290~\AA~feature is very sensitive to [\ion{Ca}{2}] emission, but not exclusively. The double peaked nature of this emission feature in the w7 model (and in 2011fe) indicates there is very little contribution from [\ion{Ca}{2}], but rather the feature is dominated by [\ion{Fe}{2}] and [\ion{Ni}{2}], while the gaussian shape of the emission from our double detonation model at this wavelength indicates it is dominated by [\ion{Ca}{2}], which is only resolvable as a doublet for low ejecta velocities. 

Recently \cite{Flors2019} performed a study of nebular spectra produced by one zone NLTE models, where the ejecta was composed of varying ratios of Fe, Ni and Co. When fitting to a sample of normal Type Ia SNe they find that they can fit the feature at 7290~\AA~in normal SNe Ia with only [\ion{Fe}{2}] and [\ion{Ni}{2}] emission, and no contribution from [\ion{Ca}{2}]. While this is likely the case for normal SNe Ia, the inferred masses of Ni and Fe indicated by the fits points to a sub-Chandrasekhar origin (by comparing to the nucleosynthetic yields of Ni and Fe for the sub-Chandrasekhar mass models in \cite{Shen2018}). However, we find that the $^{40}$Ca present in such models will produce a flux in this region, which is not seen in these normal SNe Ia \footnote{Here we define normal SNe Ia as 2011fe-like objects of magnitudes $~$-19 mag and normal velocities ($v_{Si II} \sim$ 11,000 km/s). See section \ref{sec:data} for more details.}. See section \ref{sec:Ca} for a more detailed discussion on the \cite{Shen2018} models.

\subsection{Subluminous Type Ia SNe}

The double detonation model provides a better fit to the nebular Ca emission seen in subluminous, or 91bg-like, SNe Type Ia. The bottom panel of Figure \ref{fig:w711fe} shows a model comparison to 1999by \citep{Silverman2012}, a 91bg-like SN. We choose 1999by over 1991bg because of its better signal to noise ratio, particularly for wavelengths less than 6000~\AA~where the Fe emission features are prominent. Our model naturally produces the strength and width of the [\ion{Fe}{2}] and [\ion{Fe}{3}] emission peaks as well as the strength of the [\ion{Ca}{2}] $\lambda\lambda$7291, 7323 doublet.

\cite{Mazzali91bg} investigate the nebular spectra of 1991bg using tomographic methods to determine an abundance and density profile for the ejecta. This study shows that lower central densities are required to reproduce the narrow Fe line emission features seen in this subluminous event, and these densities are consistent with both a sub-Chandrasekhar mass model and a WD merger model. However, the sub-Chandrasekhar model is ruled out due to a poor photospheric match to 1991bg. Our models, however, are a good fit for subluminous SNe Ia in the photospheric phase \citep{Polin2019} likely because we are comparing spectra for a lower mass object than the one chosen in \cite{Mazzali91bg}.

It is worth also discussing the discrepancies in this fit. Two emission features are under-produced by our models, one at $\sim$4400~\AA~and one at $\sim$9500~\AA. The 4400~\AA~emission is due to [\ion{Fe}{2}] with a small contribution from [\ion{S}{2}]. The fact that our models under produce flux at this wavelength could be due to the stable Fe arguments made for 2011fe (see section \ref{subsec:normal}). However, this feature is also under represented in the Chandrasekhar mass models produced in \cite{2011fe_nebular}. More likely this discrepancy is a result due to a limitation in the atomic data. The feature at 9500~\AA~is more puzzling. The emission in this region is primarily due to forbidden [\ion{S}{3}] emission \citep{Janos2017}. It is unclear why we under produce this emission. It is possible that this is due to too little sulfur produced at the requisite densities in the ejecta. 
While examining a sub-Chandrasekhar mass (0.9 \Msun) model, ignited as a pure central detonation with no helium on its surface, \cite{Blondin99by} successfully model the spectra of SN~1999by throughout its evolution. This model reproduces the strength of the [\ion{S}{3}] emission seen in 1999by, however also overproduces [\ion{Ca}{2}] $\lambda\lambda$7291, 7323 emission.

\subsection{SN~2018byg}

SN~2018byg (ZTF18aaqeasu) \citep{De2019} was a peculiar Type I SN found in the outskirts of its host galaxy. Features in both the light curve and spectra distinguished this as an unusual event. The light curve exhibited a rapid rise that turned out to be an excess in flux over the first few days post explosion. Spectra taken during this time show a blue continuum with some broad absorption features bluer than 5000~\AA. At peak the r-band photometry reached a maximum brightness of -18.27 $\pm$ 0.04 mag, subluminous for a Type Ia, but typical for a 91bg-like Type Ia. The spectra at peak best serve to distinguish this event as unusual. The blue part of the spectrum exhibited extreme  line blanketing, nearly extinguishing all flux for wavelengths less than 5000~\AA, and wavelengths red of this sharp cutoff exhibit absorption features from \ion{Si}{2} and a broad \ion{Ca}{2} absorption feature.  These were all features predicted by \cite{Polin2019} as ``smoking gun" signatures of a double detonation resulting from the ignition of a thick helium shell.

\cite{De2019} presented a custom model which was created following the methods of \cite{Polin2019} which showed all of these peculiarities for a 0.76 \Msun~WD with a 0.15 \Msun~helium shell. In the model the early flux excess was created by the radioactive decay of elements produced during the burning of the massive helium shell, and the extreme line blanketing was caused by those same optically thick helium ashes which reside in the outermost SN ejecta. The velocity and strength of the \ion{Ca}{2} absorption feature was well modeled, with the 0.035 \Msun~of $^{40}$Ca produced during nucleosynthesis. 

\begin{figure}
    \centering
    \includegraphics[width=\columnwidth]{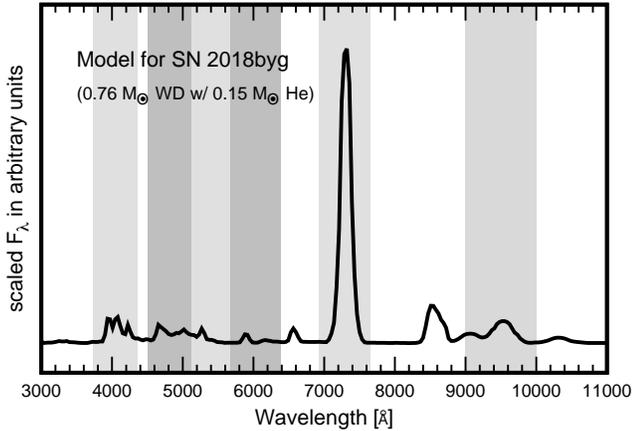}
    \caption{The \cite{Polin2019} model used to fit SN~2018byg (ZTFaaqeasu) in the photospheric phase followed through to the nebular phase. We predict that this peculiar event would fall into the regime where the nebular spectrum is dominated by [\ion{Ca}{2}] emission.} 
    \label{fig:18byg}
\end{figure}

In Figure \ref{fig:18byg} we show the consequences of running the model for SN~2018byg into the nebular phase. The total mass of the model (0.9 \Msun) places it in the [\ion{Ca}{2}] dominated regime. While no nebular spectra were taken for this event, we propose that the result would have categorized this event as Ca-rich. We predict that the next such 18byg-like event should exhibit strong [\ion{Ca}{2}] emission features in the nebular phase. 

\section{[\ion{Ca}{2}] Emission and Ca-Rich Transients}
\label{sec:Ca}
\begin{figure*}[ht]
    \centering
    \includegraphics[width=\columnwidth]{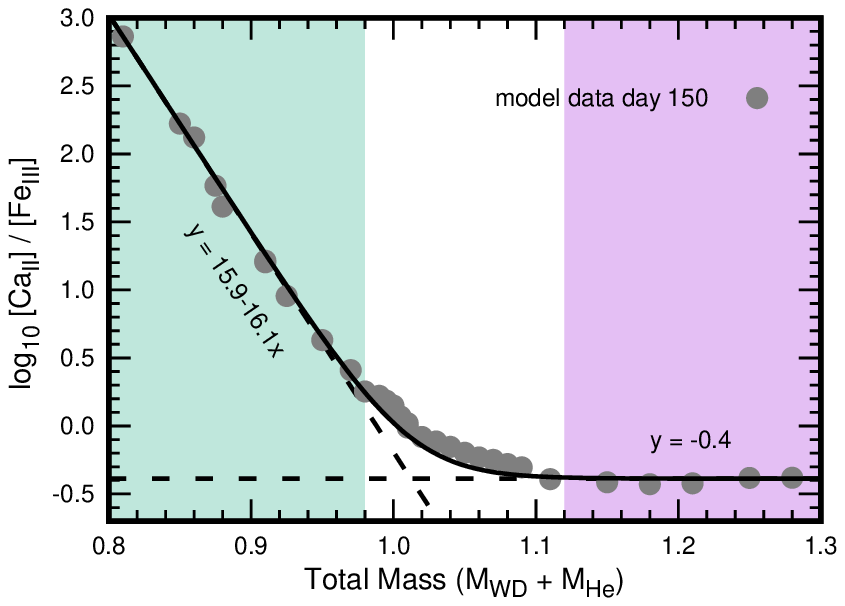}
    \includegraphics[width=\columnwidth]{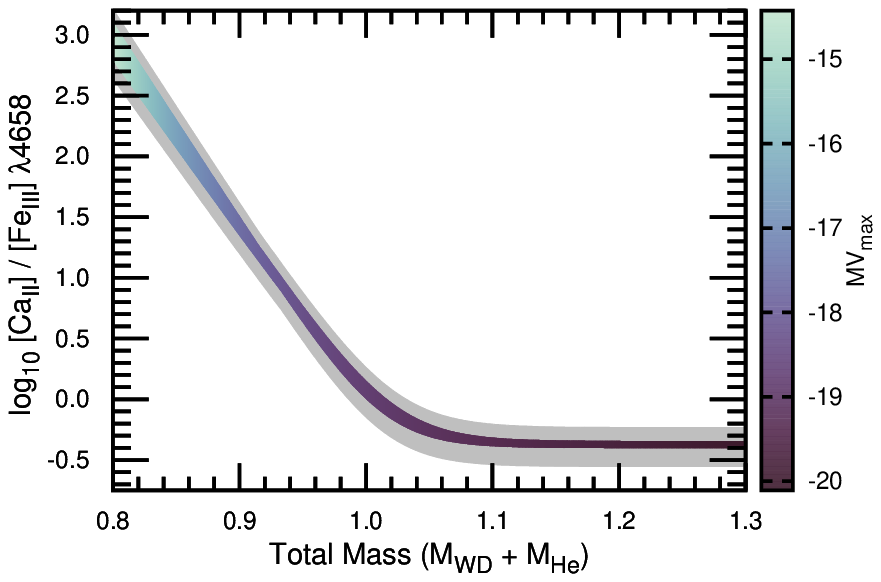}
    \caption{\textit{Left}: The relationship between [\ion{Ca}{2}]/[\ion{Fe}{3}] to the total mass of the progenitor. Grey points show data for all models at 150 days after explosion. The two exponential fits delineate two regions of parameter space: one dominated by [\ion{Ca}{2}] emission (shaded in green) and one with strong [\ion{Fe}{2}] and [\ion{Fe}{3}] emission features (shaded in purple). \\
    \textit{Right}: The results of the same fits but for all nebular times. The width of the colored line represents the variation over modeled times. The curve is bounded on the bottom by models at 150 days after explosion and on top at day 450. The grey region represents the error bars due to uncertainties in atomic data, which is reported to be $\sim$30\% in \cite{Janos2017}. Only a small variation in this ratio is measured over nebular times.} 
    \label{fig:powerlaw}
    \label{fig:CaoverFe}
\end{figure*}

\begin{figure}
    \centering
    \includegraphics[width=\columnwidth]{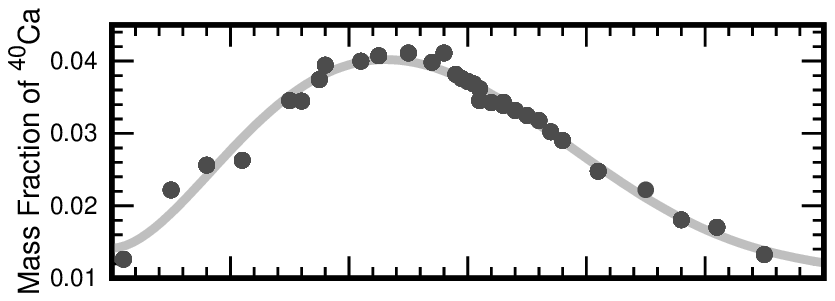}\vspace{-2.5em}\\
    \includegraphics[width=\columnwidth]{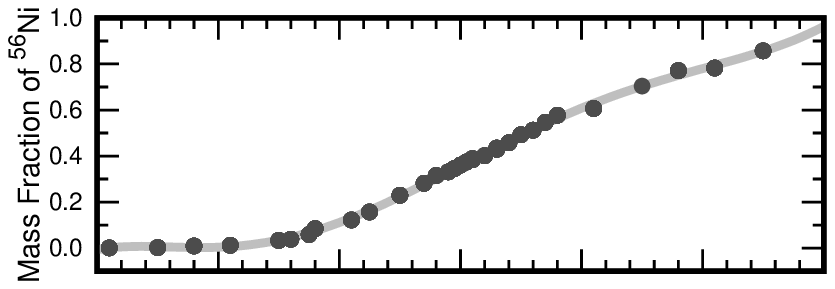}\vspace{-2.5em}\\
     \includegraphics[width=\columnwidth]{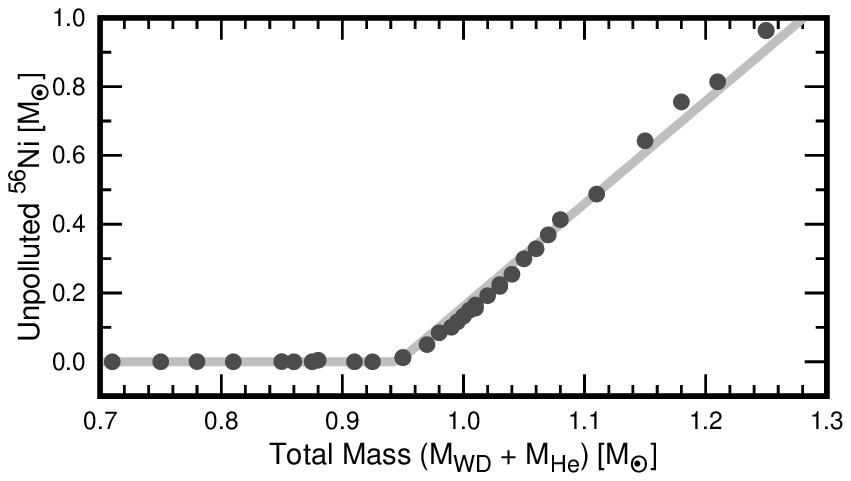}\\
    \caption{Total mass fractions of $^{40}$Ca (top) and $^{56}$Ni (middle) integrated over each ejecta model and plotted as a function of the total mass of the model. The bottom panel shows the summed mass of the \el{Ni}{56} which is unpolluted by \el{Ca}{40} at the 1\% level. Models with progenitor masses less than 0.95 \Msun~have \el{Ca}{40} mixed throughout the Ni core and appear Ca-rich in the nebular phase. Grey lines denote spline fits to the data, except in the bottom panel where the data is fit to the line $f(x) = 3.0x-2.84$ for masses greater than 0.95 \Msun.} 
    \label{fig:CaNimass}
\end{figure}

\begin{figure}
    \centering
    \includegraphics[width=\columnwidth]{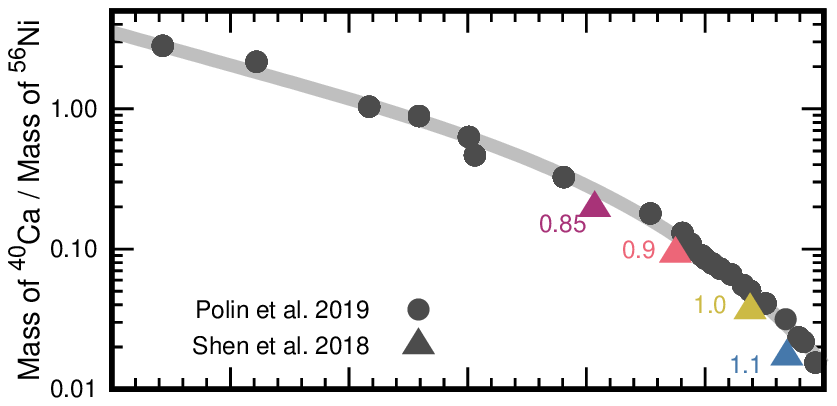}\vspace{-3em}\\
    \includegraphics[width=\columnwidth]{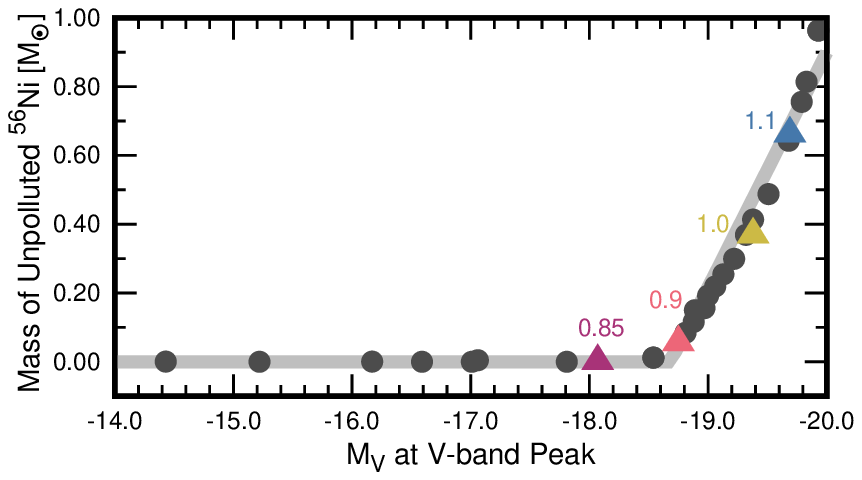}
    \caption{The ratio of the masses of $^{40}$Ca to $^{56}$Ni (top) and unpolluted $^{56}$Ni (bottom) as a function of $V$-band magnitude for the \cite{Polin2019} and \cite{Shen2018} models. The different models show good agreement in composition yields and distribution when comparing models of similar peak brightness.}
    \label{fig:MVcomp}
\end{figure}

Here we examine the striking feature in the synthesized nebular spectra: the [\ion{Ca}{2}] $\lambda\lambda$7291, 7323 doublet. Figure \ref{fig:powerlaw} shows how the ratio of [\ion{Ca}{2}]~$\lambda\lambda$7291, 7323 to [\ion{Fe}{3}]~$\lambda 4658$ emission lines (hereafter [\ion{Ca}{2}]/[\ion{Fe}{3}]) varies systematically with WD mass. The results can be fit with a sum of two exponentials, one representing a region of parameter space where the spectra are dominated by Ca emission (models with a total mass $\lesssim$ 0.98 \Msun) and one describing the Fe dominated spectra (models with a total mass $\gtrsim$ 1.1 \Msun). The transition between the two regimes occurs where the amount of intermediate mass elements produced by the progenitor is comparable to the mass of $^{56}$Ni produced. This occurs for a progenitor $\sim$1.0 \Msun~ (see Figure 2 in \cite{Polin2019}). For progenitors with masses greater than 1.0 \Msun~the gains in $^{56}$Ni are no longer exponentially increasing with the total mass and the relationship between Fe and Ca emission stays nominally constant.

The left panel of Figure \ref{fig:powerlaw} shows an example of this fitting process for all models at 150 days after explosion. The right panel of Figure \ref{fig:CaoverFe} shows the compiled fits across all nebular times with the thickness of the line being bound on top by the fit at day 450 and the bottom by day 150 post explosion. The shaded grey region represents an estimate of uncertainties in the modeled spectra due to uncertainties in atomic data \citep{Janos2017}. We show only small variations in this modeled ratio over nebular times, making this measurement a good quantity to compare to data, as some flexibility in observational epoch may be allowed.

How much to trust this prediction boils down to how confident we are about the quantities of Ca and Fe-group elements in the modeled SN ejecta, as well as their distribution throughout the ejecta. The exact strength of the [\ion{Ca}{2}] emission lines is sensitive to both the amount of \el{Ca}{40} produced in the explosion and how that calcium is distributed throughout the ejecta. [\ion{Ca}{2}] is an extremely efficient cooling line and even a very small abundance of calcium will dominate over other coolants (see for example \cite{Fransson1989} who explore this phenomenon by examining the effects of mixing calcium into oxygen regions in Type II SNe). In our double detonation models calcium is likely to dominate the cooling when it is co-produced with other coolants throughout the ejecta, as opposed to existing primarily in a localized region. In the case of our low mass progenitors there is a smaller quantity of Fe-group elements produced during burning, and the calcium region extends further into the core. Together these work to allow [\ion{Ca}{2}] emission to dominate over the Fe-group emission for low mass progenitors (roughly $M_{tot} <$ 0.9 \Msun). Figure \ref{fig:CaNimass} shows the summed mass fractions of \el{Ca}{40} and \el{Ni}{56} for all models (top two panels), as well as the amount of that \el{Ni}{56} which is uncontaminated by \el{Ca}{40} at the 1\% level (bottom panel). For progenitors with a total mass less than 0.95 \Msun~the calcium region extends fully into the nickel core, for progenitor masses greater than this the percentage of nickel which is uncontaminated by calcium increases linearly with total mass. Examining Figure \ref{fig:thinHeshellspectra} we see that this cutoff is roughly where we transition from nebular spectra which resemble SNe Ia (exhibiting Fe-group emission lines) to nebular spectra which more resemble Ca-rich transients (exhibiting [\ion{Ca}{2}] emission but lacking Fe-group lines). While the precise quantitative properties of this distribution is subject to change for multi-D simulations, which allow for asymmetries, we do expect this qualitative trend to hold.

In Figure \ref{fig:MVcomp} we compare our ejecta models to those of \cite{Shen2018} who use a sophisticated nuclear network to post-process yields for the detonations of bare sub-Chandrasekhar mass WDs (which lack helium on their surface and are ignited by a central detonation). We plot the ratio of $^{40}$Ca to $^{56}$Ni in the ejecta (top) and the mass of uncontaminated \el{Ni}{56} (bottom) vs the $V$-band magnitude at time of $V$-band peak for all of our models (grey circles) and the \cite{Shen2018} models (colored triangles). We choose $V$-band for this comparison because it is most similar across our models of the same total mass as it tends to be free from any line blanketing effects caused by the helium shell ashes, so we are free to compare them to the \cite{Shen2018} models which lack helium shells. We see that the \cite{Shen2018} models fall onto the relationship we map out relating the photospheric signatures ($M_V$) to the ejecta variables that control the [\ion{Ca}{2}] emission in the nebular phase. The masses of these models do not perfectly align (the \cite{Shen2018} 0.85 \Msun~model looks like our 0.9 \Msun~model) as mentioned in \cite{Polin2019}. However, the overall relationship is consistent. Given the masses and distribution of $^{40}$Ca and $^{56}$Ni in the \cite{Shen2018} models we expect to see this strong [\ion{Ca}{2}] emission feature even for updated nuclear networks. Due the close agreement shown in Figure \ref{fig:MVcomp} we expect to see similar nebular spectra from models in both studies that correspond to the same peak luminosity.

\subsection{Implications for Ca-Rich Transients}

Here we examine the implications of our results for Ca-rich transients. $^{40}$Ca is produced via the double detonation mechanism in quantities greater than a standard Chandrasekhar mass model for a Type Ia SN. For example the w7 model has ~0.01 \Msun~$^{40}$Ca where as our models can produce up to a factor of four times that amount. Whether or not a model would be categorized as a Ca-rich SN if observed during the nebular phase is dependent on three (related) parameters: the amount of $^{40}$Ca in the ejecta, the amount of Fe-group elements, and the relative distribution between the two. It is not sufficient to have Ca in the ejecta, in fact it is more important that the ejecta lack an Fe-group region through which to cool.

Our understanding of Ca-rich transients is limited by the small number of observed events, however current constraints made by fitting light curve shapes place ejecta mass estimates of order $\sim$0.5 \Msun~\citep{Kasliwal2012}. The question remains: how rich in Ca do these events need to be? The prototype event, SN~2005E was originally modeled with an ejecta mass of 0.3 \Msun~and 0.14 \Msun~of $^{40}$Ca in the ejecta ($X_{Ca}$=0.47) \citep{Perets2010}. More recently \cite{Dessart2015} modeled a helium shell detonation model (one where the shell burning does not cause an ignition of the underlying WD) and showed that the nebular spectra appeared Ca-rich with an ejecta mass of 0.2 \Msun~having produced 0.03 \Msun~Ca ($X_{Ca}$=0.15) and showed that the nebular spectrum continued to cool through [\ion{Ca}{2}] when that mass fraction was halved. 

While the models focused on in \cite{Polin2019} are not a good match to most Ca-rich transients in the photospheric phase, they provide important insight into the puzzle of the origin of such events. When we examine our double detonation models which would be characterized as Ca-rich (those with a total mass less than $\sim$0.9 \Msun) we see spectra that primarily cool through [\ion{Ca}{2}] emission with a significantly smaller ratio of Ca required. Figure \ref{fig:CaNimass} shows that as little as 0.02 \Msun~$^{40}$Ca (or 0.01 by mass fraction) can produce a Ca-rich event. This is an order of magnitude less than the percentages provided for Ca-rich transients in previous literature. This result may not be that surprising given that it is well established that [\ion{Ca}{2}] is a strong cooling line. For example Type II SNe can exhibit strong [\ion{Ca}{2}] emission peaks in the nebular phase when the ejecta contains very little calcium. However, those nebular spectra still contain other cooling lines, and the ratio of their [O I] to [\ion{Ca}{2}] emission lines distinguish them from the Ca-rich gap transients (See for example the Type IIP SN~2004et \citep{Jerkstrand2012}). The low mass double detonations are thermonuclear events which do qualify for the nebular Ca-rich criterion while containing very little calcium overall. We therefore support the trend to start calling these events Ca-strong rather than Ca-rich \citep{Shen2019CaStrong}.

Very low mass ($\lesssim$ 0.7 \Msun) double detonation progenitors were not explored in \cite{Polin2019}, and this low mass part of the parameter space may be a possible progenitor for some Ca-rich gap transients. This region of double detonations is worth exploring, however, NLTE radiative transport calculations would be required throughout the photospheric phase to determine if features such as helium absorption seen in many of the Ca-rich population can be explained by this model. We leave it as a future exercise to examine if very low mass double detonations (should they disrupt their underlying C/O WD) can act as potential progenitors for these Ca-rich transients.

\section{Comparison to Data}
\label{sec:data}
In this section we compare our model parameters to a set of 37 observed nebular SNe Ia, and determine how the nebular phase can further help to identify which SNe Ia arise from double detonation progenitors.

\subsection{Population of Observed Nebular SNe Ia}

We select a population of Type Ia SNe observed in the nebular phase by searching WISeREP, a public repository of SNe photometry and spectra \citep{wiserep}, and selecting all SNe Ia observed between 120 and 320 days after peak. We include all Types of SNe Ia (e.g. 91T-like, 91bg-like and peculiar) as well as normal SNe Ia. We also searched the open supernova catalog \citep{OSC} for any remaining SNe with nebular spectra from the population in \cite{Zheng2018}. The result is 37 SNe Type Ia observed in the nebular phase, all of which are plotted in Figures \ref{fig:MBwdata} and \ref{fig:siII}. See Table \ref{table:citations} for a table of all SNe and associated references.

\subsection{[\ion{Ca}{2}]/[\ion{Fe}{3}] Ratio as a Function of Magnitude}

\begin{figure}
    \centering
    \includegraphics[width=1.02\columnwidth]{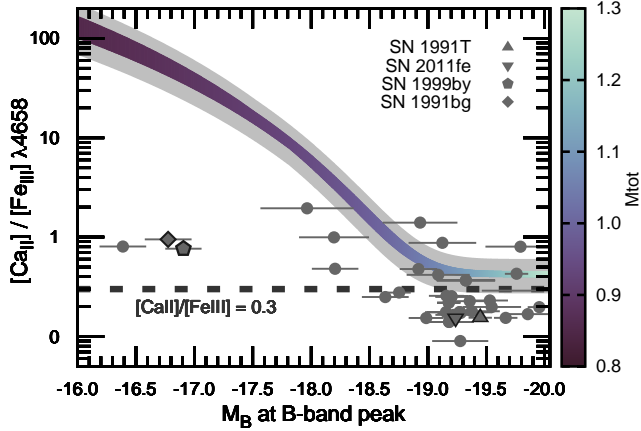}
    \caption{The modeled relationship between [\ion{Ca}{2}]/[\ion{Fe}{3}] and $M_B$ (at time of $B$-band peak) plotted with the nebular SNe data. SNe featured in Figure \ref{fig:thinHeshellspectra} are denoted with individual symbols. The modeled relationship fits a family of SNe with magnitudes brighter than -19.0 mag, but no less luminous SNe fall on the modeled region. However, we do see a rise in the strength of the [\ion{Ca}{2}] line (with respect to the [\ion{Fe}{3}] line) occurring for SNe with magnitudes less than -19.0 mag matching the predicted behavior. The dashed line represents the minimum value for [\ion{Ca}{2}]/[\ion{Fe}{3}] expected from our models while considering the grey error regions.} 
    \label{fig:MBwdata}
\end{figure}

In Figure \ref{fig:MBwdata} we compare the observed SNe data to our models by plotting the [\ion{Ca}{2}]/[\ion{Fe}{3}] ratio as a function of $B$-band magnitude at $B$-band peak. We convert between total mass of our models and the $B$-band magnitude according to the results of the thin helium shell magnitudes modeled in the photospheric phase in \cite{Polin2019}. 

Again the grey region represents the errors in atomic data from \cite{Janos2017}. The results show a trend in the increase of [\ion{Ca}{2}] emission for low luminosity events as predicted by our models. For more luminous SNe (magnitudes brighter than -19.0) we see examples of SNe with comparable and even more extreme [\ion{Ca}{2}]/[\ion{Fe}{3}] ratios than our models predict. On the low luminosity end we over predict the strength of the [\ion{Ca}{2}] line compared to the observed SNe Type Ia.

\subsection{\ion{Si}{2} Velocity Relationship: Which SNe Ia are Double Detonations?}

\begin{figure}
    \centering
    \includegraphics[width=1.02\columnwidth]{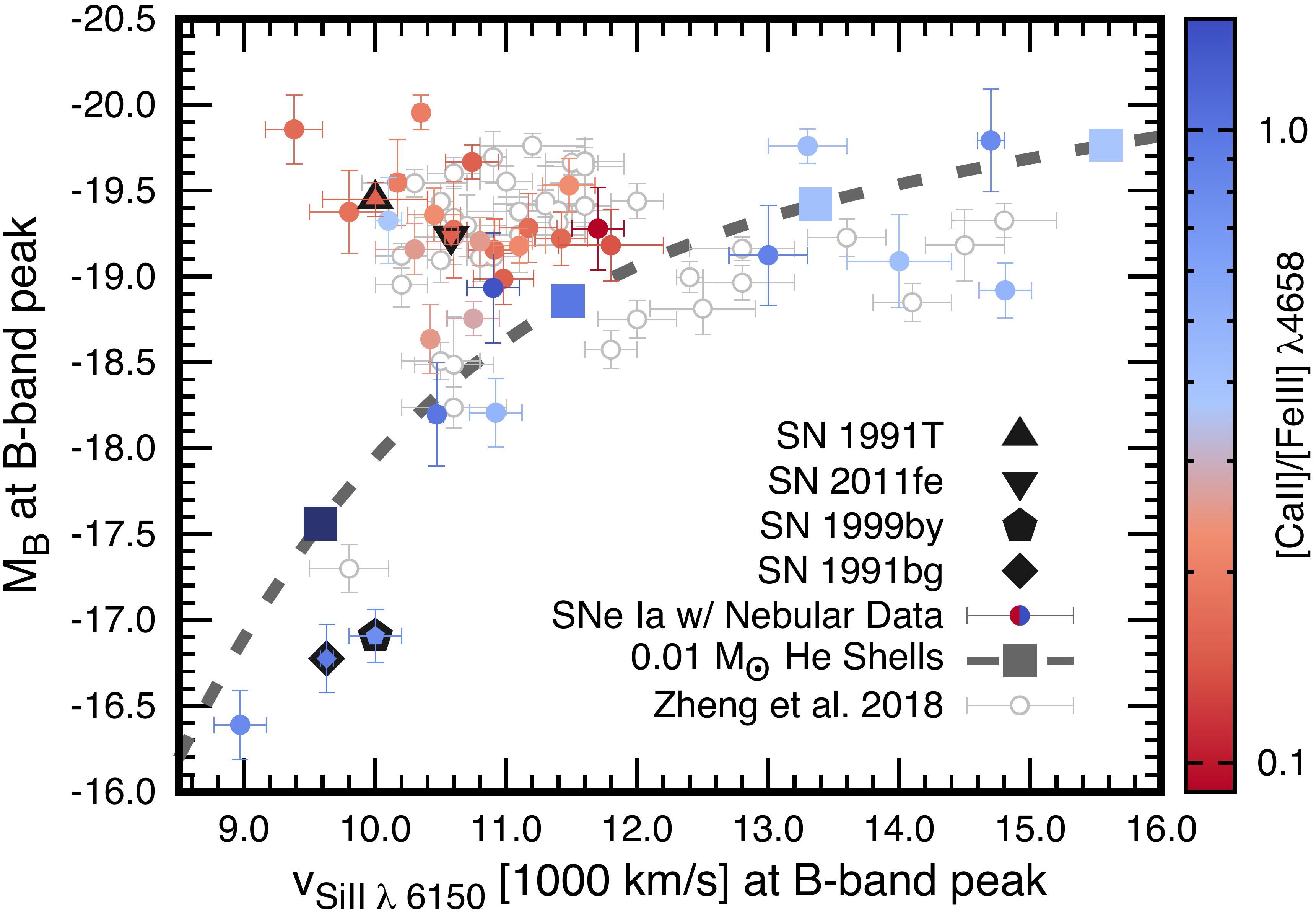}
    \caption{$B$-band magnitude vs \ion{Si}{2} velocity (both at $B$-band peak) for a observed population of SNe Type Ia. \cite{Polin2019} suggested that this relationship is evidence of two classes of SNe Ia; those that follow the modeled relationship may be identified as sub-Chandrasekhar mass progenitors and those that lie in the cluster may have a Chandrasekhar mass origin. SNe featured in Figure \ref{fig:thinHeshellspectra} are denoted with individual symbols. SN~2011fe and SN~1991T lie in the cluster while SN~1991bg and SN~199by lie along our velocity curve. All of the SNe that follow the modeled relationship have a stronger contribution of [\ion{Ca}{2}] emission than those in the cluster, supporting the hypothesis that these are the SNe Type Ia originating from a double detonation mechanism. The colorbar transition from red to blue is placed at a value of [\ion{Ca}{2}]/[\ion{Fe}{3}]=0.3, corresponding to the minimum expected value from our models.}
    \label{fig:siII}
\end{figure}

Here we examine the consistency of our nebular predictions with the population of SNe Type Ia that \cite{Polin2019} point to as the likely candidates for sub-Chandrasekhar mass double detonations. This population is identified by their relationship between \ion{Si}{2} velocity (taken from the minimum of the \ion{Si}{2} $\lambda 6150$ absorption feature) and $B$-band magnitude, both at time of $B$-band peak. \cite{Polin2019} identifies two distinct populations of SNe Type Ia. One group follows the relationship modeled by the double detonation models, and contains SNe at both high and low \ion{Si}{2} velocities. These are the SNe Type Ia that may originate from a sub-Chandrasekhar mass double detonation progenitor. The other group clusters tightly around -19.5 mag and 11,000 km/s outside of the allowed relationship for double detonations. Recently \cite{Cikota19} examined this relationship using an additional population of SNe Ia and show this trend persists in their data, although the outliers are not as easily identifiable around the Chandrasekhar mass cluster. Furthermore, when they examine spectropolarimetry measurements of the \ion{Si}{2} line polarization they yet again find a distinction between the cluster (which have lower polarization measurements) and the SNe that follow the predicted sub-Chandrasekhar mass relationship (which have higher polarization measurements).

We reproduce this plot in Figure \ref{fig:siII} with all of our nebular data colored by [\ion{Ca}{2}]/[\ion{Fe}{3}] emission. Open circles represent the \cite{Zheng2018} data for which no nebular were available. We see that this trend of outliers is further delineated by their nebular features. All of the \ion{Si}{2} velocity outliers that lie along the modeled sub-Chandrasekhar mass sequence show a [\ion{Ca}{2}]/[\ion{Fe}{3}] ratio greater than 0.3 while the majority of the SNe in the cluster show a weaker [\ion{Ca}{2}] contribution. This cutoff is the same as indicated by the minimal expected value for the ratio of [\ion{Ca}{2}]/[\ion{Fe}{3}] (see the dashed line in Figure \ref{fig:MBwdata}). The 0.01 \Msun~helium shell models are plotted as squares along a spline fit to the model data shown as a dashed grey line. From left to right the plotted models are 0.9 \Msun, 1.0 \Msun, 1.1 \Msun, and 1.2 \Msun~WDs. The 0.9 \Msun~model, having [\ion{Ca}{2}]/[\ion{Fe}{3}] $\sim$ 10 does over produce [\ion{Ca}{2}] emission when compared to any observed SNe Type Ia, however higher mass models are consistent with the data. We stress that these objects are now delineated by four different features: velocity, color \citep{Polin2019}, polarization \citep{Cikota19}, and nebular [\ion{Ca}{2}] emission. We are confident that we now distinguish two distinct classes of SNe Type Ia that differ by their physical origin.

\section{Discussion}
\label{sec:discussion}
In this study we performed a systematic survey of double detonation models in the nebular phase. We showed that these models can qualitatively reproduce subluminous SNe Ia spectra in the nebular phase, but over produce [\ion{Ca}{2}] emission compared to most normal SNe Ia. The ratio of [\ion{Ca}{2}] to [\ion{Fe}{3}] emission as a function of total mass can be used as a diagnostic to determine if a SN Ia can result from a sub-Chandrasekhar mass progenitor. When we examine this ratio compared to the peak $B$-band magnitude of our models it further supports the conclusions of \cite{Polin2019} that a population of SNe Type Ia with sub-Chandrasekhar mass double detonation progenitors can be identified by their relationship between \ion{Si}{2} velocity and $B$-band magnitude. This is now the fourth axis by which these groups differentiate from each other, which strongly indicates that two distinct classes of SNe Type Ia can be identified by this relationship.

Furthermore our study gives important insight into the progenitors of Ca-rich transients as we provide a model for a thermonuclear transient where very little Ca is needed in the SN ejecta to produce a nebular spectrum that cools predominantly through forbidden [\ion{Ca}{2}] emission. We predict that future events like SN~2018byg could be classified as Ca-rich in the nebular phase and caution the use of this diagnostic to group SNe as the photospheric spectra can be wildly disparate.

Nebular spectra are a powerful probe of the internal structure of SNe, including any asymmetries in the geometry of the ejecta. This study has been performed with 1D hydrodynamic models and we plan to perform future studies with multi-D simulations. The trends we see in the 1D models presented here are dominated by the presence and distribution of \el{Ni}{56} and \el{Ca}{40}. While the quantities of these elements produced during the explosion are subject to change in future multi-D simulations there are no multi-D physical phenomena (i.e. Rayleigh-Taylor instabilities, jets, etc.) that are lacking from the 1D simulations which would significantly alter the distribution of these elements. Therefore we do expect the qualitative trends we lay out to hold. What is compelling for a multi-D study is the ability to examine the effects of any asymmetries in the SN ejecta. It is possible for the velocity distribution to become significantly asymmetrical if the core ignition occurs off center (see for example the 2D double detonation model in \cite{Townsley2019}). It is unknown if this behavior will be uniform throughout the parameter space of possible WD and He shell mass combinations, however these asymmetries, when present, should reveal themselves in the nebular phase. 

\acknowledgments{
We would like to thank the anonymous referee for helpful comments that
improved the quality of this paper, as well as Ken Shen for useful discussions and access to the ejecta profiles from \cite{Shen2018}. We would like to thank the National Energy Research Scientific Computing Center, which is supported by the Office of Science of the U.S. Department of Energy under Contract No. DE-AC02-05CH11231, for providing staff, computational resources, and data storage for this project and the Computational HEP program in The Department of Energy's Science Office through Grant \#KA2401022. This work was supported in part by the U.S. Department of Energy, Office of Science, Office of Nuclear Physics, under contract number DE-AC02-05CH11231 and DE-SC0017616, and by a SciDAC award DE-SC0018297.  This research benefited from collaboration supported by the Gordon and Betty Moore Foundation through Grant GBMF5076.}

\software{SedoNeb \citep{Janos2017}, \\ Sedona \citep{sedona}, Castro \citep{CASTRO}}

\begin{deluxetable*}{cccccccccccccccccc}
\tablecaption{List of SNe with Nebular Spectra. \label{table:citations}}
\tabletypesize{\scriptsize}
\setlength{\tabcolsep}{4pt}
\tablehead{
\colhead{SN} & \colhead{Figures} & \colhead{Database} & \colhead{Citation(s)}
} 
\startdata 
\toprule
SNe in the \cite{Zheng2018} sample with Nebular spectra:\\
2000cx & \ref{fig:MBwdata},\ref{fig:siII}  & OSC &\cite{Silverman2012}, \cite{Li2001} \\
2001ep & \ref{fig:MBwdata},\ref{fig:siII}  & OSC  &\cite{Silverman2012} \\
2002bo & \ref{fig:MBwdata},\ref{fig:siII}  & OSC  &\cite{Blondin2012}, \cite{Matheson2008} \\
2002dj & \ref{fig:MBwdata},\ref{fig:siII}  & OSC  &\cite{Pignata2008} \\
2002er & \ref{fig:MBwdata},\ref{fig:siII}  & WISeREP  &\cite{Kotak2005} \\
2002fk & \ref{fig:MBwdata},\ref{fig:siII}  & OSC  &\cite{Hicken2009} \\
2003cg & \ref{fig:MBwdata},\ref{fig:siII}  & OSC  &\cite{Elias-Rosa2006} \\
2004dt & \ref{fig:MBwdata},\ref{fig:siII}  & OSC  &\cite{Silverman2012} \\
2004eo & \ref{fig:MBwdata},\ref{fig:siII}  & WISeREP  &\cite{Pastorello2007} \\
2005cf & \ref{fig:MBwdata},\ref{fig:siII}  & WISeREP  &\cite{Wang2009} \\
2005ki & \ref{fig:MBwdata},\ref{fig:siII}  & WISeREP  &\cite{Folatelli2013} \\
2006X & \ref{fig:MBwdata},\ref{fig:siII}  & WISeREP  &\cite{Wang2008} \\
2007af & \ref{fig:MBwdata},\ref{fig:siII}  & WISeREP  &\cite{Silverman2012} \\
2007le & \ref{fig:MBwdata},\ref{fig:siII}  & OSC  &\cite{Silverman2012} \\
\hline
Additional Nebular SNe Type Ia: \\
1986G & \ref{fig:MBwdata},\ref{fig:siII}  & WISeREP  &\cite{Cristiani1992} \\
1990N & \ref{fig:MBwdata},\ref{fig:siII}  & WISeREP  &\cite{Gomez1998} \\
1991T & \ref{fig:thinHeshellspectra},\ref{fig:MBwdata},\ref{fig:siII}  & WISeREP  &\cite{Gomez1998} \\
1991bg & \ref{fig:thinHeshellspectra},\ref{fig:MBwdata},\ref{fig:siII}  & OSC  &\cite{Turatto1996} \\
1994ae & \ref{fig:MBwdata},\ref{fig:siII}  & WISeREP  &\cite{Silverman2012} \\
1995D & \ref{fig:MBwdata},\ref{fig:siII}  & WISeREP  &\cite{Blondin2012} \\
1996X & \ref{fig:MBwdata},\ref{fig:siII}  & WISeREP  &\cite{Salvo2001} \\
1998aq & \ref{fig:MBwdata},\ref{fig:siII}  & WISeREP  &\cite{Branch2003} \\
1998bp & \ref{fig:MBwdata},\ref{fig:siII}  & WISeREP  &\cite{Silverman2012} \\
1998bu & \ref{fig:MBwdata},\ref{fig:siII}  & WISeREP  &\cite{Matheson2008} \\
1999aa & \ref{fig:MBwdata},\ref{fig:siII}  & WISeREP  &\cite{Silverman2012} \\
1999by & \ref{fig:thinHeshellspectra},\ref{fig:w711fe},\ref{fig:MBwdata},\ref{fig:siII}  & WISeREP  &\cite{Silverman2012} \\
2000E & \ref{fig:MBwdata},\ref{fig:siII}  & WISeREP  &\cite{Blondin2012} \\
2002cs & \ref{fig:MBwdata},\ref{fig:siII}  & WISeREP  &\cite{Silverman2012} \\
2002dp & \ref{fig:MBwdata},\ref{fig:siII}  & WISeREP  &\cite{Silverman2012} \\
2003du & \ref{fig:MBwdata},\ref{fig:siII}  & WISeREP  &\cite{Anupama2005} \\
2003hv & \ref{fig:MBwdata},\ref{fig:siII}  & WISeREP  &\cite{Leloudas2009} \\
2005ke & \ref{fig:MBwdata},\ref{fig:siII}  & WISeREP  &\cite{Folatelli2013} \\
2006D & \ref{fig:MBwdata},\ref{fig:siII}  & WISeREP  &\cite{Silverman2012} \\
2007if & \ref{fig:MBwdata},\ref{fig:siII}  & WISeREP  &\cite{Blondin2012} \\
2008A & \ref{fig:MBwdata},\ref{fig:siII}  & WISeREP  &\cite{McCully2014} \\
2008Q & \ref{fig:MBwdata},\ref{fig:siII}  & WISeREP  &\cite{Silverman2012} \\
2011fe & \ref{fig:thinHeshellspectra},\ref{fig:w711fe},\ref{fig:MBwdata},\ref{fig:siII}  & WISeREP  &\cite{2011fe_nebular} \\
\hline
Additional Spectra: \\
2005cz (Ca-rich transient) &\ref{fig:thinHeshellspectra} & OSC & \cite{2005cz}
\enddata
\tablecomments{The OSC refers to the Open Supernova Catalog \citep{OSC}, and the figures column refers to the figure in which the spectrum or data taken from the spectrum appear.} 
\end{deluxetable*}

\bibliographystyle{yahapj}
\bibliography{main.bib}

\end{document}